\documentclass[aps,prd,showpacs]{revtex4}
\usepackage{amsmath,amssymb}
\usepackage{graphicx,color,epsfig}

\begin{document} 

\title{Perfect fluid spheres with cosmological constant}

\author{Christian G.~B\"ohmer}
\email{c.boehmer@ucl.ac.uk}
\affiliation{Department of Mathematics, University College London,
             Gower Street, London, WC1E 6BT, UK}

\author{Gyula Fodor}
\email{gfodor@rmki.kfki.hu}
\affiliation{KFKI Research Institute for Particle and Nuclear Physics,
             H-1525 Budapest, P.O.Box 49, Hungary}

\date{\today}

\begin{abstract}
  We examine static perfect fluid spheres in the presence of a
  cosmological constant.  Due to the cosmological constant, new
  classes of exact matter solutions are found.  One class of solutions
  requires the Nariai metric in the vacuum region. Another class
  generalizes the Einstein static universe such that neither its
  energy density nor its pressure is constant throughout the
  spacetime. Using analytical techniques we derive conditions
  depending on the equation of state to locate the vanishing pressure
  surface. This surface can in general be located in regions, where
  going outwards, the area of the spheres associated with the group of
  spherical symmetry is decreasing. We use numerical methods to
  integrate the field equations for realistic equations of state and
  find consistent results.
\end{abstract}

\pacs{04.20.Jb, 04.40.Dg, 04.20.-q}

\maketitle

\section{Introduction}

Static and spherically symmetric perfect fluid solutions have always been
a rich source of investigation in classical general relativity ever since the
pioneering work of Schwarzschild in 1916. He solved the field equations 
for the interior region by assuming a perfect fluid of constant energy 
density, and also for the outside vacuum region, the famous Schwarzschild
solutions.  

The interior solution has the geometry of a three-sphere, which was
noticed as early as 1919 by Weyl~\cite{Weyl:1919}.  The pressure of
the Schwarzschild interior solution always vanishes before the equator
of the three-sphere, therefore, this geometrical picture is not of
great importance. This situation chances significantly in the presence
of the cosmological constant~\cite{Boehmer:2002gg,Boehmer:2003uz}.  In
some special cases the interior metric with cosmological constant was
studied earlier, see e.g.~\cite{Xu:1986ka,Stuchlik:2000}.  However,
the complete analysis was only
completed~\cite{Boehmer:2002gg,Boehmer:2003uz} 85 years after Weyl
noted the interesting geometrical structure of these solutions: With
$\Lambda$ the pressure can vanish exactly at the equator of the
three-sphere in which case one has to join on the Nariai
metric~\cite{Nariai1,Nariai2} as the exterior vacuum metric.  We note
that although the spatial geometry of the interior Schwarzschild is a
three sphere, the four-metric is not homogeneous. There is a center of
symmetry in the fluid region, surrounded with concentric spheres
defined by the orbits of the group generating spherical symmetry. The
spacetime metric associate an induced metric and thereby an area to
each group orbit. Going outwards from the center this area is
increasing until one reaches the equator, the sphere with the maximal
area.  It is furthermore possible that the pressure vanishes in a
region where the area of the group orbits is decreasing.

Lastly, the matter can occupy the whole three-sphere having two regular centers.
This generalizes the Einstein static 
universe~\cite{Einstein:1917ce,Boehmer:2002gg,Boehmer:2003uz}, 
see~\cite{Ibrahim:1976} for early results in that direction. The Einstein universe
also emerges as a special case (vanishing expansion rate and vanishing vorticity) 
when considering homogeneous shear-free perfect fluids~\cite{Collins:1988}. In recent
years it has been generalized in various different theories, like
brane world models~\cite{Gergely:2001tn}, Einstein-Cartan theory~\cite{Boehmer:2003iv},
in modified gravity theories~\cite{Boehmer:2007tr} or Loop Quantum 
Gravity~\cite{Parisi:2007kv}.

In this paper we are analyzing systematically the effects of a positive cosmological 
constant on perfect fluid spheres. We generalize known exact solutions of the field 
equations with cosmological term and discuss their new properties. The principal
result of the analysis is the increase of the radial sizes of matter spheres. This 
naturally lead to questions regarding the physical picture applicable to these solutions. 
Analytical and numerical techniques are used to obtain a consistent picture of the 
underlying physics.

A positive cosmological constant can be regarded as an external force
pulling matter apart. Therefore, a `large' positive cosmological constant 
can increase the radius of a known perfect fluid solution such that
it occupies more than just `half' of the three-space. The effects of
the actual cosmological constant are very small and therefore one may
ask why the investigation of solutions with `large' values of the
cosmological term is beneficial. This can be answered from two points
of view. From a mathematical point of view we deepen our understanding
about exact solutions of Einstein's field equations and the influence
of the additional parameter $\Lambda$.  However, also from a physical
point of view this study can be justified easily. In the bag model of
hadrons~\cite{Chodos:1974je,Chodos:1974pn,DeGrand:1975cf} the bag 
is stabilized by a term of the form $B g_{ab}$, which has the same form
as a cosmological constant, though its numerical value is considerably
bigger $B_{e}^{1/4}\approx 8.91 {\rm MeV}$, whereas we have
$\Lambda^{1/4}\approx 1.78 \times 10^{-9} {\rm MeV}$. Hence, the
key physical motivation is the possibility that effects within stellar
models can be effectively described by a term that looks like a
cosmological constant, which can indeed have large effects.  Also in
the context of Boson stars~\cite{Kaup:1968,Ruffini:1969qy} the matter
energy-momentum tensor contains a part proportional to the metric
which can be read as an effective cosmological term. Within the context
of Loop Quantum Gravity it has recently been shown~\cite{Boehmer:2007wi}
that quantum gravity effects can be effectively of the form of a 
cosmological constant.

Apart from an equation of state relating the density $\rho$ to the
pressure $p$, a spherically symmetric perfect fluid has to satisfy
only one condition, the pressure isotropy condition, requiring the
equality of the radial and angular directional pressures. 
Since this condition does not involve the cosmological
constant $\Lambda$, any solution with $\Lambda=0$ having an equation
of state $f(\rho,p)=0$ can also be interpreted as a $\Lambda\neq 0$
cosmological solution with equation of state
$f(\rho+\Lambda,p-\Lambda)=0$.  It is one of the main purposes of the
present paper to re-investigate known perfect fluid solutions with
cosmological constant and to analyze whether these solutions for some
special cosmological constant require the Nariai metric as the
exterior vacuum spacetime. We also try to construct solutions with two
regular centers, i.e.~constructing more general Einstein static
universes, having neither constant energy density nor constant pressure.  

Since the static and spherically symmetric field equations with more realistic
equations of state in general cannot be integrated analytically, we
also use numerical methods to study perfect fluid spheres. In
particular we will show that the effect of a `large' cosmological
constant on polytropic perfect fluids is such that the matter is
pulled sufficiently apart so that it occupies much of the
three-space. The same result is also found by considering the stiff
matter equation of state and also the Hagedorn equation of
state. However, realistic equations of state seem not to allow the
presence of a second regular center. It is important to distinguish
between coordinate and physical effects. For instance, one cannot
numerically integrate the constant density solutions for large
cosmological constants if the original radius is used as a
variable. The code breaks down at the equator of the
three-sphere. This is a pure coordinate effect, since the coordinate
system does not cover the whole spacetime. In order to distinguish
coordinate and physical effects, also the Riemann curvature
tensor is considered.

Various other effects of the cosmological constant have been studied
in the past, like the dynamical instability of perfect fluid 
spheres~\cite{Boehmer:2005kk,Hledik:2007tr}, possibly detectable effects 
of the cosmological constant within our solar 
system~\cite{Iorio:2005vw,Sereno:2006re,Jetzer:2006gn,Kagramanova:2006ax}, 
and effects within astrophysical
structures~\cite{Balaguera-Antolinez:2004sg,Balaguera-Antolinez:2005wg,Balaguera-Antolinez:2005yy,Balaguera-Antolinez:2007mx,BalagueraAntolinez:2007sf}. Recently, the bending of light 
with $\Lambda$ has been discussed in~\cite{Rindler:2007zz,Sereno:2007rm}.

The paper is outlined in the following manner: In Section~\ref{whitt} we analytically
study the effect of the cosmological constant on the Whittaker and Tolman $IV$ solution,
and also discuss the matching of the matter and the vacuum solution. In Section~\ref{sec:analytic}
we derive conditions on the equation of state and the cosmological constant to characterize 
the different possible solutions. In Section~\ref{sec:num} we numerically integrate the
field equations for realistic equations of state and find results consistent with our 
analytical results. We summarize and conclude our work in the final Section~\ref{sec:concl}

\section{The Whittaker and Tolman solutions}
\label{whitt}

In the present section the Whittaker solution~\cite{Whittaker:1968}
and the Tolman $IV$ solution~\cite{Tolman:1939jz} are recalled.  In both
cases we firstly introduce a third angular coordinate $\alpha$, so that the
coordinate system covers the complete three-space and not just `half'
of it like the usual radial coordinate $r$. Secondly, we introduce the
cosmological constant in the solutions and analyze its effect (the
`external' force due to $\Lambda$) on those solutions.  We also
explicitly show, how to join the interior and the exterior solution
through the vanishing pressure surface.

\subsection{The Whittaker solution}
\label{subsec:whitt}

The Whittaker solution is characterized by the relation $\rho + 3p =\rho_0$ 
between the energy density and the pressure, where $\rho_0$
is a positive constant. Similarly to the constant density case this
condition allows one to write down the solution to Einstein's field
equations in terms of elementary functions. The metric of the
Whittaker solution in the original Schwarzschild coordinates
reads~\cite{Whittaker:1968}
\begin{multline}
      ds^2 = -b\left[1+B-\frac{B}{ar}\sqrt{1-a^2r^2}\arcsin(ar)
             \right]dt^2 \\
             +\left[(1+B)(1-a^2r^2)-\frac{B}{ar}(1-a^2r^2)^{3/2}
             \arcsin(ar)\right]^{-1}dr^2 +r^2d\Omega^2 ,
      \label{whitt1}
\end{multline}
where $a$, $b$ and $B$ are constants.  The special importance of the
Whittaker solution lies in the fact that it is the non-rotating static
limit of the Wahlquist solution~\cite{Wahlquist:1968}, the most
important rotating perfect fluid exact solution. The parameter
$\kappa$ of the Wahlquist solution is related to the parameter $B$ of
the Whittaker metric by $B=1/\kappa^2$.  Since the change of the
parameter $n$ corresponds merely to a rescaling of the coordinate $t$,
we set $b=1$.  After introducing a new radial coordinate, the third
angle $\alpha$, by $r=(1/a)\sin\alpha$, metric~(\ref{whitt1})
simplifies to
\begin{align}
      ds^2 &= -f dt^2
            +\frac{1}{a^2}\left(\frac{d\alpha^2}{f}
            +\sin^2\negmedspace\alpha\,d\Omega^2\right), 
      \label{whitt2} \\
      f &= 1+B(1-\alpha\cot\alpha) .
      \label{whitt3}
\end{align}
Although the introduced radial coordinate $\alpha$ is very similar to
the third angle of the ellipsoid used for the interior Schwarzschild
solution, the spatial metric is not ellipsoidal in this case because
of the non constant nature of the metric component $g_{\alpha\alpha}$.
On the other hand, it should be emphasized that the $r={\rm const.}$
hypersurfaces are round two-spheres. Therefore, the topology of the
three-space essentially depends on the function $g_{\alpha\alpha}$. In
the above mentioned constant density case, the three-space is in fact
a three-sphere.  When we discuss next the modified Tolman $IV$ solution
the three-space will be ellipsoidal. The introduction of the
new radial coordinate $\alpha$ is important in all cases where there
is a group orbit with maximum area, since in these cases the usual
radial coordinate only covers the region up the this maximum orbit.
Henceforth we will refer to the coordinate $\alpha$ as the third
angle.

It is possible to express the constant in the equation of state $\rho
+ 3p =\rho_0$ in terms of the constants in the metric and the
cosmological constant by
\begin{align}
\rho_0=\frac{a^2B+\Lambda}{4\pi}  \label{rho0} .
\end{align}
Positivity of the pressure and density implies $\rho_0>0$, which we
require from now on, by assuming
\begin{align}
\Lambda>-a^2B \label{rho0pos}.
\end{align}
If there is a spherical surface where the solution is matched to a
Schwarzschild-de Sitter (or Schwarzschild anti-de Sitter) exterior
region, then $p$ must go to zero at the surface, and the fluid density
becomes $\rho_0$ there.

The pressure and energy density of the Whittaker solution are given by
\begin{align}
      p &= \frac{\rho_0}{2} - \frac{a^2f}{8\pi} ,
      \label{whitt_p} \\
      \rho &= \frac{3a^2f}{8\pi}-\frac{\rho_0}{2} .
      \label{whitt_rho}
\end{align}
The central pressure and central energy density can be derived from
equations~(\ref{whitt_p}) and~(\ref{whitt_rho}) by noting that
$\lim_{\alpha\rightarrow 0}f=1$ and read
\begin{align}
      p_c = \frac{\rho_0}{2} - \frac{a^2}{8\pi} ,\qquad
      \rho_c = \frac{3a^2}{8\pi}-\frac{\rho_0}{2} .
      \label{whitt_cen}
\end{align}
Requiring both to be positive we get
$\frac{4\pi}{3}\rho_0<a^2<4\pi\rho_0$.  Since at the center $\frac{df}{d\alpha}=0$ and
$\frac{d^2f}{d\alpha^2}=\frac{2B}{3}$, the pressure is maximal at the
center if and only if $B>0$. Since in the $B<0$ case there is no zero
pressure surface, we assume $B>0$ in the present discussion.

The equator, where the area of the spheres of symmetry is maximal, is
located at $\alpha=\pi/2$. There $f=1+B$ and for the equatorial
pressure $p_{eq}$ and density $\rho_{eq}$ we get
\begin{align}
      p_{eq} = \frac{1}{8\pi}(\Lambda-a^2) ,\qquad
      \rho_{eq} = \frac{3}{8\pi}(a^2-\Lambda)+\rho_0 .
      \label{whitt_pi2}
\end{align}

Let us now choose the cosmological constant such that the pressure
vanishes at the equator, $p_{eq}=0$ in~(\ref{whitt_pi2}), which yields
\begin{align}
      \Lambda = a^2 =: \Lambda_N.
      \label{whitt5}
\end{align}
Hence for solutions which satisfy the relation $\Lambda < \Lambda_N$,
the pressure vanishes before the equator, where the group orbits
are still increasing.

One should check whether the choice $\Lambda=\Lambda_N$ is compatible
with the positivity of pressure and energy density at the center, and
also with the positivity of energy density at the equator (otherwise
these solutions would not be physical).  Positivity of energy density
at the surface is ensured by (\ref{rho0pos}), which now takes the form
$B>-1$.  Using $\Lambda=a^2$ in (\ref{whitt_cen}) and (\ref{rho0}) we
get
\begin{align}
      p_c &= \frac{\Lambda_{N}}{8\pi}B > 0 ,
      \label{whitt6} \\
      \rho_c &= \frac{\Lambda_{N}}{8\pi}(2-B) > 0 .
      \label{whitt7} 
\end{align}
All the three conditions are satisfied if $0<B<2$. 

For larger cosmological constants, i.e.~for $\Lambda > \Lambda_N$, the
pressure vanishes after the equator of the ellipsoid, where the group
orbits are decreasing.  Since the pressure function $p\rightarrow
-\infty$ as $\alpha\rightarrow\pi$ there always exists a zero pressure
surface at which one joins on the Schwarzschild-de Sitter metric as
an exterior vacuum spacetime. Therefore
the Whittaker solution cannot have a second regular center.  Since the
derivative of the function $f$ is positive for $0<\alpha<\pi$ in the
case $B>0$, the function $f$ remains positive, consequently the
coordinate system and the metric remains regular in the whole fluid
region.

As already outlined in the introduction we will explicitly show that
the choice $\Lambda=\Lambda_N$ necessitates the Nariai metric as the
exterior spacetime. Since the pressure in this case vanishes at the
maximum of the area of the group orbits the exterior spacetime must
have constant area spheres of symmetry, which excludes the
Schwarzschild-de Sitter or anti-de Sitter spacetimes. The only other
static and spherically symmetric vacuum spacetime with cosmological
constant is indeed the Nariai spacetime, and its group orbits have
constant area.

\subsection{Matching procedure}
\label{subsec:match}

Here we review the necessary conditions for matching a spherically
symmetric static perfect fluid solution to an exterior vacuum
region. We write the metric in both regions in the form
\begin{align}
      ds^2=-e^{2\nu}dt^2+\frac{1}{y^2}dr^2+R^2(d\theta^2+\sin^2\theta d\phi^2),
      \label{matchds}
\end{align}
where $\nu$, $y$ and $R$ are functions of the coordinate $r$.  A
metric written in this form have a center where $R=0$, and this center
is regular, i.e. free of conical singularities, if the area of small
spheres is proportional to their radius square, with the appropriate
proportionality factor, $\frac{dR}{dr}=\pm\frac{1}{y}$. The regularity
of the four-metric also requires a finite value for $\nu$ and
$\frac{d\nu}{dr}=0$ at the central point.

We assume that the matching is performed along the hypersurface
described by $r=r_s$.  The Darmois-Israel matching
conditions~\cite{Darmois:1927,Israel:1966rt} essentially state that
the induced metric and the extrinsic curvature have to agree on the
hypersurfaces used for joining the two solutions.  The outward
pointing normal vector to the symmetry surfaces has the components
$n^a=(0,y,0,0)$. The induced metric $h_{ab}$ can be expressed using
the spacetime metric as $h_{ab}=g_{ab}-n_a n_b$, while the extrinsic
curvature can be calculated as $K_{ab}=h_{a}^{\ c}\nabla_c n_b$, where
$\nabla_c$ denotes the covariant derivative associated to the
spacetime metric $g_{ab}$. Using the coordinate system
$(t,\theta,\phi)$ on the matching surface, the induced metric $h_{ab}$
has the form
\begin{align}
      h_{ab}=
      \begin{pmatrix}
            -e^{2\nu} & 0 & 0 \\
            0 & R^2 & 0 \\
            0 & 0 & R^2\sin^2\theta
      \end{pmatrix}, \label{hab}
\end{align}
while the components of the extrinsic curvature are
\begin{align}
      K_{ab}=
      \begin{pmatrix}
            -y e^{2\nu}\frac{d\nu}{dr} & 0 & 0 \\
            0 & y R\frac{dR}{dr} & 0 \\
            0 & 0 & y R\frac{dR}{dr}\sin^2\theta 
      \end{pmatrix}. 
      \label{Kab}
\end{align}
From~(\ref{hab}) it is apparent that the induced metric agrees if and
only if the value of $R$ and $\nu$ agrees on the matching surfaces.
Then the extrinsic curvatures are matched appropriately if and only if
$y \frac{d\nu}{dr}$ and $y \frac{dR}{dr}$ agrees. The agreement of $R$
has the obvious physical meaning of equal area matching spheres, while
the equality of $\nu$ can always be ensured by appropriate rescaling
the time coordinate in the interior fluid domain. If we use Gauss
coordinates in both domain then $y=1$ and the matching of the
extrinsic curvature is equivalent to the continuity of the first
derivative of $\nu$ and $R$.  However, in general, it is also possible
to give coordinate system invariant meaning to these conditions.

The invariant mass function in spherically symmetric cosmological
spacetimes can be defined as
\begin{equation}
      m=\frac{R}{2}(1-g^{ab}R_{,a}R_{,b}) - \frac{\Lambda}{6}R^3.
\end{equation}
For vanishing cosmological constant this gives back the usual mass
definition given in \cite{Zannias:1990}.
For the metric form~(\ref{matchds}) we get
\begin{equation}
      m=\frac{R}{2}\left[1-y^2\left(\frac{dR}{dr}\right)^2\right]
      - \frac{\Lambda}{6}R^3.
      \label{anmass}
\end{equation}
From this we can see that if $m$ agrees on the two matching surfaces of
equal area then $y \frac{dR}{dr}$ must also be the same. The other
invariant quantity is the pressure at the matching surface, which for
the metric (\ref{matchds}) takes the form
\begin{align}
      p=\frac{y^2}{R}\frac{dR}{dr}\left(2\frac{d\nu}{dr}
      +\frac{1}{R}\frac{dR}{dr}\right)+\Lambda.
\end{align}
It is apparent that if $R$ and $y \frac{dR}{dr}$ both agree on the
matching surfaces and $\frac{dR}{dr}\neq0$ then $y \frac{d\nu}{dr}$
will be the same if and only if the pressures are the
same. Consequently, if $\frac{dR}{dr}\neq0$, the matching of two
static perfect fluid solutions can be done at two chosen spherical
surfaces if and only if the surfaces has the same area, the mass
function has the same value, and the pressures agrees as well.
Obviously, if the exterior domain is a vacuum, then the fluid pressure
at the surface must vanish. It is interesting that in case of a Nariai
exterior $\frac{dR}{dr}=0$ and the $p=0$ condition is not enough to
ensure the continuity of $y \frac{dR}{dr}$.

The quantity $y \frac{d\nu}{dr}$ is closely related to the
acceleration of static non-rotating observers staying at constant
radius $r$. In the coordinate system $x^a=(t,r,\theta,\phi)$ used in
\eqref{matchds} these observers have the velocity vector
$v^a=(e^{-\nu},0,0,0)$. The only non-vanishing component of their
acceleration vector $a^a=u^b\nabla_b u^a$ is
$a^r=y^2\frac{d\nu}{dr}$. The norm of the acceleration is
$|a|=\sqrt{a^a a_a}=y\left|\frac{d\nu}{dr}\right|$. This shows that
apart from a possible signature change the continuity of the magnitude
of the acceleration implies the continuity of $y \frac{d\nu}{dr}$ in
the matching condition.

\subsection{Joining interior and exterior solution} 

\begin{figure}[!h]
\centering\epsfig{figure=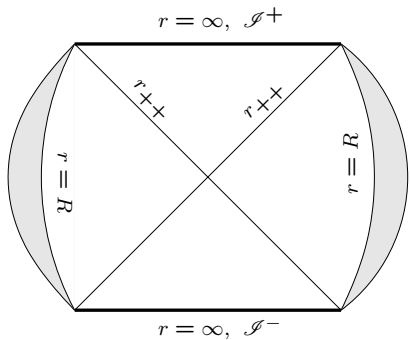,width=0.3\linewidth} 
\caption{
Penrose-Carter diagram with two stellar objects separated by a
Schwarzschild-de Sitter vacuum domain. The radii $R$ of the stellar
objects is between the radii of the black hole and cosmological
horizons. Since the group orbits are increasing up to $R$ the vacuum
part contains the cosmological event horizon $r_{++}$.}
\label{fig:Penrosea}
\end{figure}

For cosmological constants satisfying $\Lambda < \Lambda_N$ the
area of the group orbits at the $p=0$ surface is increasing and we 
join the Schwarzschild-de Sitter (or Schwarzschild anti-de Sitter for
$\Lambda<0$) metric on as the 
exterior vacuum spacetime. Since the cosmological constant is fixed by 
the specific solution it remains to choose the mass appropriately. 
In the Schwarzschild area coordinate $R$ the mass is defined by 
$M=\int_0^{R_s} 4\pi R^2 \rho(R)dR$, where $R_s$ is defined by $p(R_s)=0$. 
By using Gauss coordinates relative to the $r=\mbox{const.}$ 
hypersurfaces the metric is $C^1$ at the boundary. 
If the energy density is non-vanishing at the boundary
this cannot be improved. After placing one object in the Schwarzschild-de
Sitter spacetime, it still contains an infinite series of singularities.
However, by placing a second object appropriately in that spacetime, it 
is possible to construct a singularity-free spacetime, see 
Fig.~\ref{fig:Penrosea}. This possibility has been discussed earlier
in greater detail in~\cite{Boehmer:2002gg,Boehmer:2003uz}.

For $\Lambda = \Lambda_N$ we explicitely show the matching of the interior 
perfect fluid spacetime with the exterior Nariai spacetime. We follow the 
generic discussion of matching two static and spherically symmetric 
regions presented in the previous subsection. We read off and compare
at the matching surface the corresponding functions $\nu$, $y$ and $R$
in the general form of the line element \eqref{matchds}.

Recall that the static form of the Nariai metric is given by
\cite{Nariai2}
\begin{align}
       ds^2 = -\cos^2\chi\, dt^2
              +\frac{1}{\Lambda}\Bigl(d\chi^2 + d\Omega^2\Bigr).
       \label{nariai1}
\end{align}
We note that this form of the metric is not homogeneous, since the
static observers described by constant $(\chi,\theta,\phi)$ are not
equivalent. The magnitude of their acceleration is
$|a|=\sqrt{\Lambda}\tan\chi$. Somewhat surprisingly, this acceleration
is towards the $\chi=0$ surface, since the non-vanishing component of
their acceleration, given by $a^r=-\Lambda\tan\chi$, is negative for
$\chi>0$.

The metric functions of the Nariai spacetime in the coordinate system
\eqref{matchds} are given by
\begin{align}
      e^{2\nu_N}=\cos^2\chi, \qquad y_N^2=\Lambda, \qquad
      R_N^2=\frac{1}{\Lambda} \ .
\end{align}
The corresponding functions in the Whittaker fluid region are
\begin{align}
      e^{2\nu_W}=c^2 f, \qquad y_W^2=a^2 f, \qquad 
      R_W^2=\frac{\sin^2\alpha}{a^2}, 
\end{align}
where $f$ is given by \eqref{whitt3}. The agreement of the induced
metric, i.e.~the matching of $R$ and $\nu$ implies
\begin{align}
      \frac{1}{\Lambda}=\frac{\sin^2\alpha}{a^2}, \qquad
      \cos^2\chi=c^2 f.
\end{align}
The condition $y_N\frac{dR_N}{d\chi}=y_W\frac{dR_W}{d\alpha}$ implies
$\alpha=\pi/2$ for the matching surface in the fluid region. From this
it follows that $\Lambda=a^2$, which is just the condition of
vanishing pressure at the equator of the Whittaker solution. The
remaining matching condition
$y_N\frac{d\nu_N}{d\chi}=y_W\frac{d\nu_W}{d\alpha}$ yields
\begin{align}
      \tan\chi = -\frac{\pi B}{4\sqrt{1+B}}
      =\frac{\pi}{4}\left(1-\frac{4\pi\rho_0}{\Lambda}\right)
      \sqrt{\frac{\Lambda}{4\pi\rho_0}}.
\end{align}

\begin{figure}[!h]
\centering\epsfig{figure=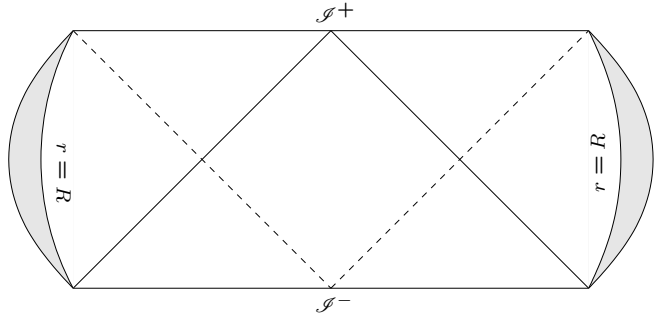,width=0.48\linewidth} 
\caption{Penrose-Carter diagram with two stellar objects having radii $R$ which 
         require the Nariai spacetime to be the vacuum part of the global solution.
         The solid and dashed lines represent the future and past event
         horizon, respectively.}
\label{fig:Penrosec}
\end{figure}

So if Gauss coordinates are used, i.e.~$y=1$, we explicitly showed
the matching of the interior and the exterior metric to be of degree
$C^1$ ($\nu$, $R$, $\nu'$ and $R'$ all agree on the zero pressure
surface). This differentiability condition cannot be improved when the
equation of state of the fluid constrains the energy density at the
boundary to be positive. In this case the energy-momentum tensor jumps 
at the boundary and the metric is at most $C^1$. 
Figure~\ref{fig:Penrosec} shows the Penrose diagram of this spacetime.

\begin{figure}[!h]
\centering\epsfig{figure=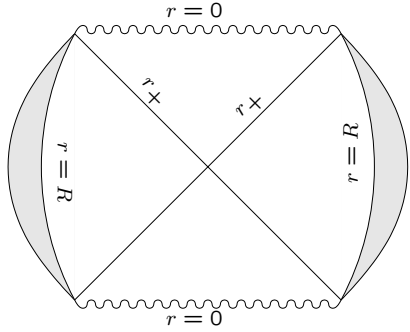,width=0.3\linewidth} 
\caption{Penrose-Carter diagram with two stellar objects connected by a 
Schwarzschild-de Sitter vacuum domain. Since the group orbits are decreasing
at the matching surface the vacuum part contains the black hole
horizon and the $r=0$ singularity.}
\label{fig:Penroseb}
\end{figure}

Lastly, for cosmological constants satisfying $\Lambda > \Lambda_N$ the
group orbits at the $p=0$ surfaces are {\em decreasing} and we join
the Schwarzschild-de Sitter metric on as the exterior vacuum spacetime.
Since the cosmological constant is fixed by the specific solution
it remains to choose the mass appropriately, see the discussion above. 
By using Gauss coordinates the metric is $C^1$ at the boundary. If 
the energy density is non-vanishing at the boundary this cannot be 
improved. It is important to note that the vacuum part of this
spacetime contains the singularity at the origin $r=0$ and that the
matter occupies the
'outer' region of the spacetime, see Penrose diagram~\ref{fig:Penroseb},
where a second object was inserted in the spacetime to remove the 
infinite sequence of singularities present in the Schwarzschild-de Sitter
diagram.

\subsection{Cosmological Tolman solutions}
\label{tolman}

In principle, the above discussion can now be repeated for all the 127 candidate
solutions presented in~\cite{Delgaty:1998uy} (of which only 60\% are isotropic 
and regular at the center). For all those one could check whether the inclusion
of the cosmological constant can pull the matter up to or beyond the equator
of the corresponding ellipsoid. We will, however, only repeat this analysis for
the Tolman $IV$ solution. This particular choice is motivated by the simplicity
of the solution which also allows to analytically express its equation of state.

We now take a fresh look at the Tolman $IV$ solution~\cite{Tolman:1939jz}.
Its metric reads
\begin{align}
      ds^2 = -B^2\bigl(1+r^2/A^2)dt^2
             +\frac{1+2r^2/A^2}{(1-r^2/R^2)(1+r^2/A^2)}dr^2
             +r^2 d\Omega^2 ,
      \label{tolman1}
\end{align}
where $A$ and $R$ are positive constants. Introducing the third angle
by $r=R\sin\alpha$ yields
\begin{align}
      ds^2 = -B^2\bigl(1+(R^2/A^2)\sin^2\negmedspace\alpha)dt^2
             +R^2 \Bigl[ \frac{1+2(R^2/A^2)\sin^2\negmedspace\alpha}
              {1+(R^2/A^2)\sin^2\negmedspace\alpha}d\alpha^2
             +\sin^2\negmedspace\alpha d\Omega^2 \Bigr] .
      \label{tolman2}
\end{align}
Pressure and energy density respectively are given by
\begin{align}
      8\pi p(\alpha)&=\frac{R^2-3R^2\sin^2\negmedspace\alpha-A^2}
                            {R^2(A^2+2R^2\sin^2\negmedspace\alpha)}+\Lambda ,
      \label{tolmanp}\\
      8\pi \rho(\alpha)&=\frac{R^2+3R^2\sin^2\negmedspace\alpha+3A^2}
                               {R^2(A^2+2R^2\sin^2\negmedspace\alpha)}+
                         \frac{2A^2\cos^2\negmedspace\alpha}
                               {(A^2+2R^2\sin^2\negmedspace\alpha)^2}-\Lambda .
      \label{tolmanr}
\end{align}
Since all quantities depend on $\alpha$ only through $\sin^2\alpha$ and
$\cos^2\alpha$, the solution is symmetric to the equator $\alpha=\pi/2$. 

Eliminating the variable $\alpha$ from Eqs.~(\ref{tolmanp}) and~(\ref{tolmanr}) leads
to the following equation of state
\begin{align}
      \rho = c_0 + c_1 p + c_2 p^2 ,
      \label{eom:tolman}
\end{align}
where the three constants $c_i$ are given by
\begin{align}
      c_0 = \frac{(3-2\Lambda R^2)(R^2+A^2(2-\Lambda R^2))}{4\pi R^2(A^2+2R^2)} ,\qquad
      c_1 = \frac{2R^2+A^2(13-8\Lambda R^2)}{A^2+2R^2} ,\qquad
      c_2 = \frac{32\pi A^2 R^2}{A^2+2R^2} .
\end{align}

As in the previous discussion let us now compute the
value of the pressure and density at the first center $(\alpha=0)$, at
the equator of the ellipsoid $(\alpha=\pi/2)$ and at the second
possible center $(\alpha=\pi)$, which yields
\begin{alignat}{2}
      8\pi p(0) &= \frac{1}{A^2}-\frac{1}{R^2} + \Lambda , &\qquad
      8\pi \rho(0) &= \frac{3}{A^2}+\frac{3}{R^2} - \Lambda, 
      \label{tolman10}\\
      8\pi p(\pi/2) &= -\frac{1}{R^2} + \Lambda , &\qquad
      8\pi \rho(\pi/2) &= \frac{3A^2+4R^2}{R^2(A^2+2R^2)} -\Lambda ,
      \label{tolman11}\\
      8\pi p(\pi) &= \frac{1}{A^2}-\frac{1}{R^2} + \Lambda , &\qquad
      8\pi \rho(\pi) &= \frac{3}{A^2}+\frac{3}{R^2} - \Lambda .
      \label{tolman12}
\end{alignat}
Similarly to the Whittaker case there exists a cosmological constant such
that the pressure vanishes at the equator
\begin{align}
      \Lambda = \frac{1}{R^2} =: \Lambda_N,
\end{align}
where the energy density is positive. In this case one has to match
the Nariai metric. For smaller cosmological constants,
i.e.~$\Lambda<\Lambda_N$, the pressure vanishes before the equator and
one has to join on the Schwarzschild-de Sitter (or Schwarzschild 
anti-de Sitter) metric as the exterior
vacuum spacetime. However, the pressure cannot vanish after the
equator, as can be seen from the mirror symmetry to the equator of the
solution. Therefore, solutions with $\Lambda > \Lambda_N$ have two 
centers. One easily verifies that both centers are regular by checking 
that the derivatives of pressure and energy density vanish at both
centers. Hence, as a side result we already found a new generalization
of the Einstein static universe. 

The Einstein universe is characterized by its constant energy energy and 
constant pressure (originally Einstein assumed a pressure-less universe) 
throughout the three-sphere. By a generalization of the Einstein static universe
we mean a globally regular solution of the field equations with cosmological
constant where the spatial part of the metric is a closed three-space and where 
either the energy density or the pressure, or both are varying.

It is expected that other known perfect fluid solutions, e.g.~those given in
Ref.~\cite{Delgaty:1998uy}, will show similar properties.  Therefore,
following the above procedure, we can explicitely show the existence of
a wide class of generalized Einstein static universes and an even wider 
class of static and spherically symmetric perfect fluid solutions, for 
which the pressure vanishes in regions where the group orbits are decreasing.

\section{Analytic considerations}
\label{sec:analytic}

In the previous sections we analyzed perfect fluid solutions
which may extend through the equator of the ellipsoid that describes the
global geometry of the spatial hypersurfaces. The present section
supplements the explicit and later numerical results by presenting
some general statements. It should also be mentioned that the 
existence and uniqueness of perfect fluid solutions was proved
in~\cite{Rendall:1991hg}. The restrictions on the equations of
state could be weakened in Refs.~\cite{Baumgarte:1993,Mars:1996gd}.
The existence and uniqueness proof of~\cite{Rendall:1991hg} could be
extended to include cosmological constants satisfying $\Lambda<4\pi\rho(p=0)$
in~\cite{Boehmer:2002gg,Boehmer:2004nu}. Let us now consider the 
static and spherically symmetric line element in Gauss coordinates 
relative to the $r={\rm const}.$ hypersurfaces
\begin{align}
      ds^2 = -e^{2\nu(r)}dt^2 + dr^2 + R^2(r)d\Omega^2,
      \label{an1}
\end{align}
the resulting field equations $G_{ab} + \Lambda g_{ab} = 8\pi T_{ab}$ 
are given by 
\begin{align}
      \frac{1-R'^2-2RR''}{R^2}-\Lambda = 8\pi \rho,
      \label{an2} \\
      \frac{R'^2-1+2RR'\nu'}{R^2} + \Lambda = 8\pi p,
      \label{an3} \\
      \frac{\nu' R' +R(\nu'^2 +\nu'')+R''}{R} +\Lambda = 8\pi p,
      \label{an4}
\end{align}
that imply the conservation of the energy-momentum tensor
\begin{align}
      p' + \nu' (p+\rho) = 0,
      \label{an5}
\end{align}
where the prime denotes differentiation with respect to $r$.
The maximum of the area of the group orbits is located
at the maximum of the function $R(r)$, which means $R'(r_m)=0$,
where $r_m$ is the location of the maximum. We henceforth
assume that such a maximum exists (To be precise we only assume
that the function $R(r)$ has an extremum and it will turn out
that this extremum is a maximum if we require the energy density
to be positive at $r_m$. For $r=r_m$ the mass definition~(\ref{anmass}) 
implies $1-2m(r_m)/R(r_m)-\Lambda/3 R(r_m)^3=0$). 

Note that mass~(\ref{anmass}) and energy density $\rho$ are related by
\begin{align} 
      \label{mprime}
      \frac{dm}{dR}=\frac{m'}{R'} = 4\pi \rho R^2.
\end{align} 

Eliminating the function $\nu'$ from the first two field
equations~(\ref{an2})--(\ref{an3}) and the conservation 
equation~(\ref{an5}) yields the Tolman-Oppenheimer-Volkoff 
(TOV)~\cite{Tolman:1939jz,Oppenheimer:1939ne} equation
\begin{align}
      \frac{dp}{dR} = -R \frac{(p+\rho)(4\pi p + m/R^3 -\Lambda/3)}
      {1-\frac{2m}{R}-\frac{\Lambda}{3}R^3},
      \label{tov}
\end{align}
where we used that $dp/dr=(dp/dR) R'$. At the maximum $r_m$ 
the TOV equation is ill-defined since the denominator tends to zero.
However, this is not a physical singularity, as can easily be seen
by considering the derivative of the second field equation~(\ref{an3}) 
evaluated at $R'=0$ which reads
\begin{align}
      8\pi p'(r_m) = 2 \nu'(r_m) \frac{R''(r_m)}{R(r_m)},
      \label{an10}
\end{align}
and moreover all Riemann tensor components~(\ref{riemann}) are well
defined at $r=r_m$. Furthermore one can the express energy density
plus the pressure function in terms of the Riemann tensor,
\begin{align}
      4\pi (\rho+p) = R^{r\theta}{}_{r\theta}-R^{\theta t}{}_{\theta t}
                    = R^{r\phi}{}_{r\phi}-R^{\phi t}{}_{\phi t},
      \label{an11}
\end{align}
so that this sum is well defined if the spacetime is non-singular.
Let us furthermore evaluate the field equations at $r_m$ which yields
\begin{align}
      \frac{1-2R(r_m)R''(r_m)}{R^2(r_m)}-\Lambda = 8\pi \rho(r_m),
      \label{an12} \\
      \frac{-1}{R^2(r_m)} + \Lambda = 8\pi p(r_m),
      \label{an13} \\
      \frac{R(r_m)(\nu'(r_m)^2 +\nu''(r_m))+R''(r_m)}{R(r_m)} 
      +\Lambda = 8\pi p(r_m). \label{an14}
\end{align}
Next, from equation~(\ref{an13}) we find 
that the pressure is positive at the equator if the cosmological
constant is large enough, this means if
\begin{align}
      \Lambda > \frac{1}{R^2(r_m)} =: \Lambda_N.
      \label{an15}
\end{align}
For the special case $\Lambda=\Lambda_N$ the pressure vanishes at
the equator, i.e.~the maximum of the area of the group orbits.
Putting this particular value of the cosmological constant into
the first field equations yields
\begin{align}
      \frac{1-2R(r_m)R''(r_m)}{R^2(r_m)}-\Lambda_N 
      = -2\frac{R''(r_m)}{R(r_m)}= 8\pi \rho(r_m),
      \label{an16}
\end{align}
from which we conclude that $R''(r_m) < 0$ to have a physically
meaningful perfect fluid solution. Indeed, this condition simply
states that the equator is a local maximum of the group orbits'
area. This fact was not assumed explicitely and is a direct 
consequence of assuming physical solutions: positivity of the 
energy density.

The above analysis clarifies under which conditions static and 
spherically symmetric perfect fluid ellipsoids may extend through
the equator and have a zero pressure surface in regions where the
area of the group orbits is decreasing. 
However, condition~(\ref{an15})
depends on the function $R(r)$ and therefore essentially depends 
on the solution. Given an equation of state, a central pressure 
and a cosmological constant, such that the pressure (and by the
equation of state the energy density) is decreasing near the 
center, the above considerations are not sufficient to decide
whether the pressure vanishes before the equator or not.
Therefore one should find a condition depending solely on the
equation of state and on the initial conditions (pressure at the 
center and the cosmological constant) so that one controls the
location of the zero pressure surface.

The TOV equation~(\ref{tov}) seems to be ill-defined at $r_m$,
since its denominator vanishes. On the other hand, the field
equations imply that the spacetime is regular where $R'(r_m)=0$.
Therefore we can conclude that the limit $\lim_{r\rightarrow r_m} p'(r)$
must exist, so that we write $\lim_{r\rightarrow r_m} p'(r)=a$.
Existence of $p'(r_m)$ can be put back into~(\ref{tov}) and yields
\begin{align}
      a = \frac{p(r_m)+\rho(r_m)}{R(r_m)^2} \lim_{r\rightarrow r_m}
      \frac{(4\pi p(r) R(r)^3 + m(r) -\Lambda/3 R(r)^3)}{R'(r)}.
      \label{a1}
\end{align} 
Since $\lim_{r\rightarrow r_m} R'(r)=0$, the numerator also must vanish
as $r\rightarrow r_m$,
\begin{align}
      \lim_{r\rightarrow r_m}(4\pi p(r)+ m(r)/R(r)^3 -\Lambda/3)=
      4\pi p(r_m)+ m(r_m)/R(r_m)^3 -\Lambda/3 = 0,
      \label{a2}
\end{align}
which can be written more conveniently (for the present purpose)
\begin{align}
      4\pi p(r_m) = \frac{1}{R(r_m)^3}
\left(\frac{\Lambda}{3}R(r_m)^3-m(r_m) \right).
      \label{a3}
\end{align}
Before exploiting the latter equation~(\ref{a3}), we note that it is
easy to show that the pressure in the TOV equation~(\ref{tov}) is 
decreasing near the center if
\begin{align}
      \Lambda < 4\pi\rho(p_c) + 12\pi p_c, 
      \label{a3a}
\end{align}
a condition that only depends on the initial values and the equation
of state.

According to equation~(\ref{a3}) the signature of the pressure at the
equator is determined by the signature of the quantity
\begin{align}
	\frac{\Lambda}{3}R_m^3- m(R_m)= 
        \int_0^{R_m} [\Lambda-4\pi\rho(R)]R^2 dR,
        \label{a4}
\end{align}
where we used the definition of mass, see~(\ref{mprime}). 
Since pressure and by a monotonic equation of state also the energy 
density are decreasing functions, $\rho(p=p_c)\geq\rho(r_m)$, and a 
sufficient condition to satisfy the inequality  $p(r_m) > 0$ is
\begin{align}
      4\pi\rho(p=p_c) < \Lambda.
      \label{a8}
\end{align}
Hence, if we the cosmological term is large enough, compared to the
central density, then the pressure does not vanish before the equator 
of the interior spacetime.

Next we find a necessary condition for a positive pressure at the
equator. Let us assume $p(r_m)\geq0$. Then the integral of
$\Lambda-4\pi\rho(R)$ is non negative. Since going outwards $p$ and 
$\rho$ are monotonically decreasing it follows that 
$\Lambda-4\pi\rho(R_m)\geq 0$. But because the zero pressure surface is 
after the equator, by the monotonicity condition we have 
$\rho(R_m)\geq\rho(p=0)$, and consequently $\Lambda\geq 4\pi\rho(p=0)$. 
On the other hand, this means that if
\begin{align}
      \Lambda < 4\pi\rho(p=0)
      \label{a5}
\end{align}  
then necessarily $p(r_m)<0$.
This condition is in agreement with previous results, see 
e.g.~\cite{Boehmer:2002gg,Boehmer:2003uz}. Hence, if the given equation 
of state and the cosmological constant satisfy the condition~(\ref{a5}),
then the pressure vanishes before the equator of the ellipsoid.

Next, let us assume that the pressure vanishes at the maximum of
the area of the group orbits, so that the equator is also the
zero pressure surface. In this case, as we already showed in the
previous sections, one has to join on the Nariai metric.
Putting $p(r_m)=p(r_b)=0$ into~(\ref{a3}) leads to
\begin{align} 
      m(r_b) = M = \frac{\Lambda}{3}R_b^3,\qquad
      \int_0^{R_b} 4\pi\rho(R)dR = \frac{\Lambda}{3} R_b^3
      \label{a6}
\end{align}
which relates the mass of the solution to the cosmological constant.
Unfortunately this condition cannot be written in a form such that the
equation of state suffices to choose the initial values for such
a solution.

These three observations can be summarized as follows
\begin{align*}
      \Lambda < 4\pi\rho(p=0)\qquad &\qquad 
      {\rm pressure\ vanishes\ before\ the\ equator,}\\
      4\pi\rho(p=0)< \Lambda <4\pi\rho(p=p_c) &\qquad 
      {\rm no\ analytical\ control,}\\
      4\pi\rho(p=p_c) < \Lambda\qquad &\qquad 
      {\rm pressure\ can\ vanish\ only\ after\ the\ equator,}
\end{align*}
which only depend on the equation of state and the initial conditions.
However, these conditions only yield quite general conclusions. They
do not suffice to decide whether the solution can have a second regular 
center, i.e.~perfect fluid solutions which occupy the whole ellipsoid. 
From equation~(\ref{a3a}) we can conclude that the pressure is increasing
near the first regular center if we assume $\Lambda >4\pi\rho(p_c)+12\pi p_c$,
however, we cannot control the further behavior of the solution and
may obtain a singular solution where the pressure diverges, or a
solution with a second center having a conical singularity. 

Let us furthermore discuss the consequences of having a regular center.
Regularity of the solution at the center in particular fixes some of the
coefficients in the power series expansion of the function 
$R(r)$, see e.g.~\cite{Fodor:2000gu}, which are given by
\begin{align}
      R(r_c) = 0, \qquad R'(r_c) = 1, \qquad R''(r_c) = 0.
      \label{a9}
\end{align}
However, the part $R'''(r_c)$ is also determined by the initial conditions,
as can be seen from the first field equation~(\ref{an2})
\begin{align}
      8\pi\rho(p_c) + \Lambda &= \lim_{r\rightarrow r_c} \frac{1-R'^2}{R^2}
      - \lim_{r\rightarrow r_c} \frac{2R''}{R}.
      \label{a10}
\end{align}
After applying the rule of L'Hopital and using the relations~(\ref{a9})
we arrive at
\begin{align}
      8\pi\rho(p_c) + \Lambda = - 3 R'''(r_c),
      \label{a11}
\end{align}
so that also the third derivative is fixed by the initial conditions.
In case there exists a second regular center $r_{c_2}$ we have
\begin{align}
      R(r_{c_2}) = 0, \qquad R'(r_{c_2}) = -1, \qquad R''(r_{c_2}) = 0
      \label{a12}
\end{align}
so that $R'''(r_{c_2})$ is given by
\begin{align}
      8\pi\rho(p_{c_2}) + \Lambda = 3 R'''(r_{c_2}).
      \label{a13}
\end{align}
The conditions~(\ref{a9}), (\ref{a11}) and also~(\ref{a12}), (\ref{a13}) 
must be satisfied independently by any solution admitting two regular
centers. While one prescribes initial conditions at the first center,
the regularity of the second center is by no means warranted since it
actually depends on the solution of the function $R$ together with the
equation of state that also enter~(\ref{a13}).

In the next section we will analyze the field equations numerically for given 
equations of state. It will turn out that none of the equations of state considered
allows a second regular center. 
Therefore, all solutions that have an increasing pressure near the
first center will have either a divergent pressure, or a second center
with a conical singularity, and are hence not of physical interest in
general.

\section{Numerical considerations}
\label{sec:num}

In sections~\ref{subsec:whitt} and~\ref{tolman} we generalized two 
known solutions to include the cosmological constant. For large cosmological
constant we found that new properties arise like the possibility of having
a second regular center. In the previous analytic section we presented some 
arguments in favor of the existence of solutions with cosmological constant 
that may occupy more than `half' the three space. This 
result can also be read in the flowing way: For sufficiently large cosmological 
constants the pressure cannot vanish before the equator of the three space,
so that some new physical properties may arise. However, it was also argued
that in general a second regular center does not exist.

It is the aim of the present section to solve the field equations~(\ref{an2})--(\ref{an4}) 
numerically for a given equation of state $\rho=\rho(p)$. In particular we are interested 
in solutions where the pressure vanishes after the group orbit's maximum of the three 
space and in solutions that possibly have two regular centers. We concentrate on these 
two classes of solutions, since the others are well known already.

\begin{figure}[!h]
\noindent
\begin{minipage}[h]{.48\linewidth}
\centering\epsfig{figure=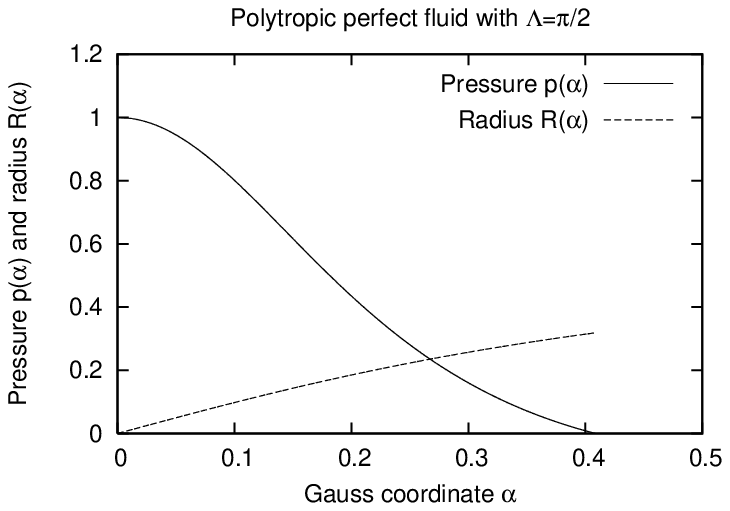,width=\linewidth} 
\end{minipage}\hfill
\begin{minipage}[h]{.48\linewidth}
\centering\epsfig{figure=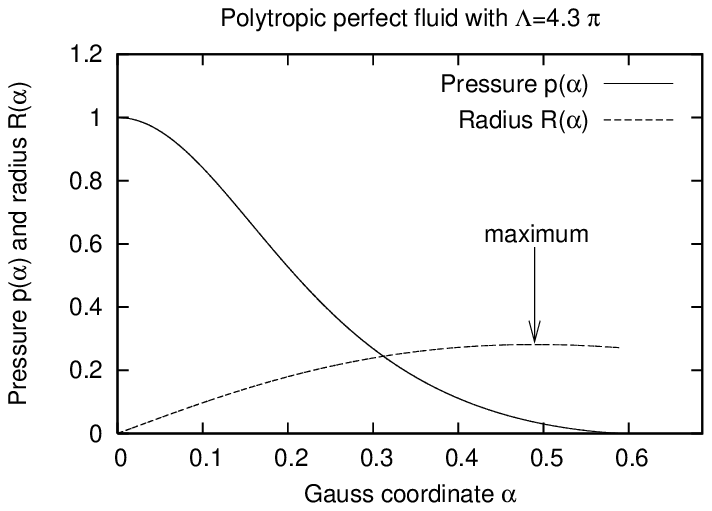,width=\linewidth} 
\end{minipage}
\caption{Pressure function and radius for the polytropic equations of state, with
         $K=1$, $n=3$ and $\rho_b=0.5$. Initial conditions are $p_c=1.0$ ($\rho_c=1.5$)
         and $\Lambda=\pi/2$ (left), $\Lambda=4.3\pi$ (right).}
\label{fig:poly}
\end{figure}

The most natural starting point for the numerical analysis are the polytropic 
equations of state
\begin{align}
      p(\rho) = K \rho^{\frac{n+1}{n}}, \qquad
      \rho(p) = \Bigl(\frac{p}{K}\Bigr)^{\frac{n}{n+1}},
      \label{num1}
\end{align}
where $K$ is some constant and $n$ is the polytropic index.
In the Newtonian case stellar models are finite if $1 < n < 5$ and do not have a 
finite radius for $n \geq 5$, where also $\Lambda=0$ is assumed. We slightly modify 
the polytropic equations of state, in order to allow a non-vanishing boundary density 
$\rho_b=\rho(p=0)$. Hence, we consider the following equation of state
\begin{align}
      \rho(p) = \Bigl(\frac{p}{K}\Bigr)^{\frac{n}{n+1}} + \rho_b,
      \label{num2}
\end{align}
where the boundary density is a new free parameter that we must specify.
We chose $K=1$, the polytropic index $n=3$ and $\rho_b=0.5$. For two
different cosmological constant (`small' and `large') we obtain the two
following solutions, Fig.~\ref{fig:poly}a and Fig.~\ref{fig:poly}b.

As expected, for small cosmological constant, see Fig.~\ref{fig:poly}a, the pressure
vanishes before the maximum of the group orbits. At the vanishing pressure surface
one can join on the Schwarzschild-de Sitter metric $C^1$ as the exterior spacetime, 
with the methods described in subsection~\ref{subsec:match}. These solutions are
represented by the Carter-Penrose diagram~\ref{fig:Penrosea} discussed earlier.
For large cosmological constant however, see Fig~\ref{fig:poly}b, the pressure vanishes 
after the maximum, in a region where the group orbits are decreasing. This would
necessarily yield a global solution represented by the Carter-Penrose 
diagram~\ref{fig:Penroseb}, where the exterior spacetime contains the singularity.
Since the numerical solutions vary smoothly in the cosmological constant, it is
evident that a fine tuned cosmological constant can be chosen such that the pressure
vanishes exactly at the maximum of the group orbits, in which case one has to join the
Nariai metric as the exterior spacetime.

\begin{figure}[h!]
\noindent
\begin{minipage}[h]{.48\linewidth}
\centering\epsfig{figure=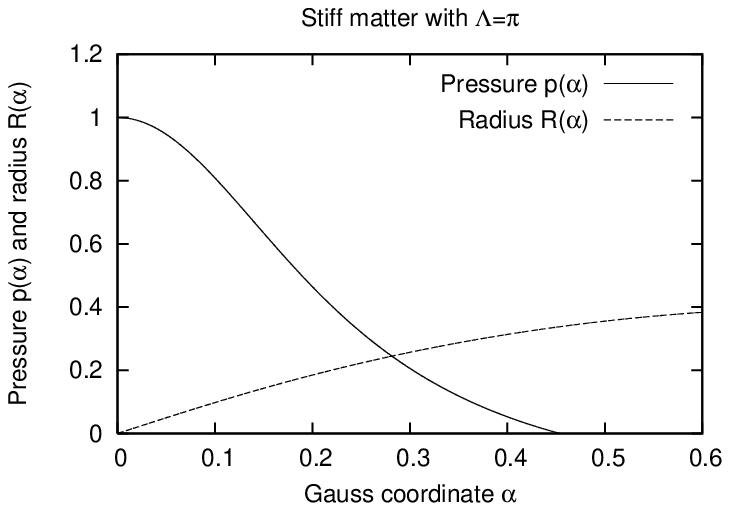,width=\linewidth} 
\end{minipage}\hfill
\begin{minipage}[h]{.48\linewidth}
\centering\epsfig{figure=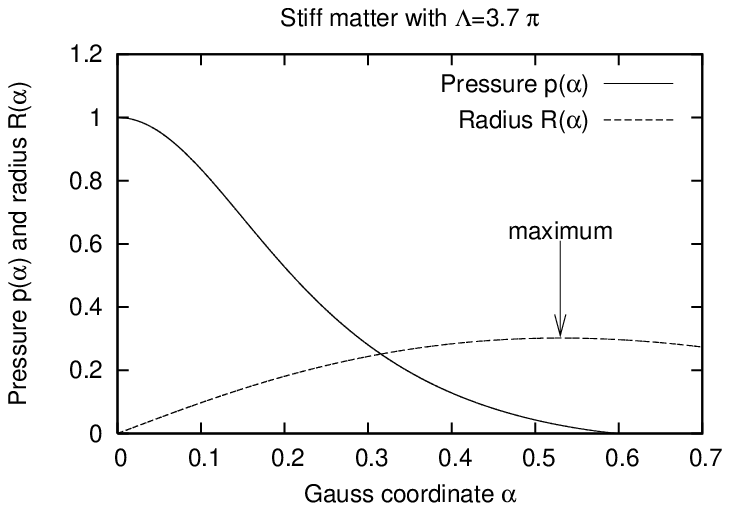,width=\linewidth} 
\end{minipage}
\caption{Pressure function and radius for the stiff matter equation of state, with
         $\rho_b=0.5$. Initial conditions are $p_c=1.0$ ($\rho_c=1.5$)
         and $\Lambda=\pi$ (left), $\Lambda=3.7\pi$ (right).}
\label{fig:stiffA}
\end{figure}

Next, let us analyze the solutions for the stiff matter equation of state
(the $n\rightarrow\infty$ limit of the polytropic equation of state)
\begin{align}
      \rho(p) = p + \rho_b,
\end{align}
which we, as before, supplemented by a boundary density term $\rho_b$. For the stiff 
matter case we again take two different values of the cosmological constant. The 
results are similar to those discussed already, see Fig.~\ref{fig:stiffA}a and 
Fig.~\ref{fig:stiffA}b.

\begin{figure}[h!]
\noindent
\begin{minipage}[h]{.48\linewidth}
\centering\epsfig{figure=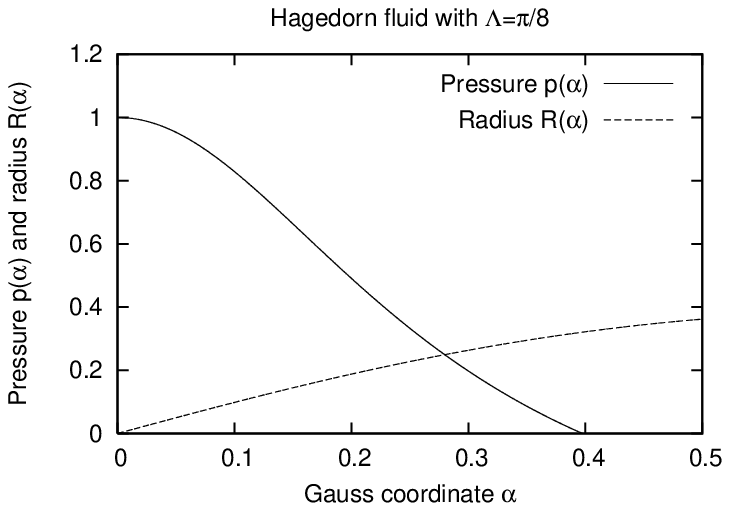,width=\linewidth} 
\end{minipage}\hfill
\begin{minipage}[h]{.48\linewidth}
\centering\epsfig{figure=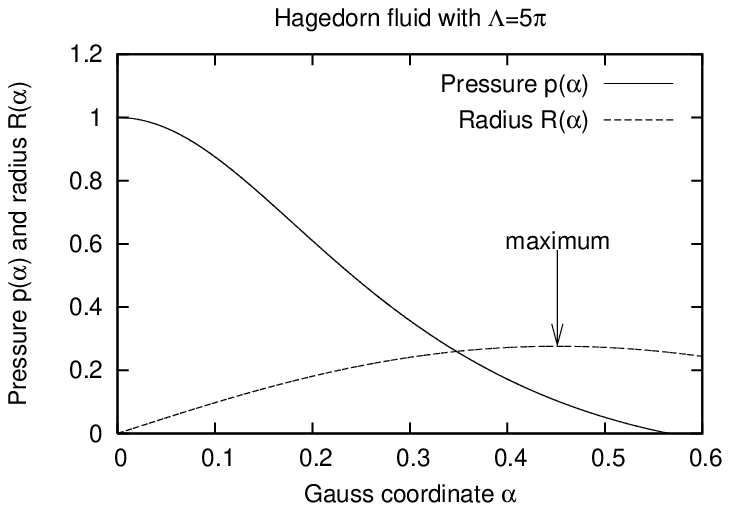,width=\linewidth} 
\end{minipage}
\caption{Pressure function and radius for the Hagedorn equation of state, with
         $\rho^{\ast}=2.0$. Initial conditions are $p_c=1.0$ ($\rho_c\approx 1.21$)
         and $\Lambda=\pi/8$ (left), $\Lambda=5\pi$ (right).}
\label{fig:hage}
\end{figure}

Finally let us consider the Hagedorn equation of state
\begin{align}
      \rho(p) = \rho^{\ast}\exp\bigl(\frac{p}{\rho^{\ast}}-1\bigr),
\end{align}
where the free parameter $\rho^{\ast}$ is related to the boundary density by
$\rho_b = \rho(p=0) = \rho^{\ast}/e$. As in the previously discussed cases, we find
that for `small' cosmological constants the pressure vanishes before the maximum
of the group orbits, whereas `large' values of the cosmological constant allow 
the pressure to vanish after the maximum, see Fig.~\ref{fig:hage}a and 
Fig.~\ref{fig:hage}b.

Apart from the incompressible perfect fluid case we could not find numerically 
any other configuration with a second regular center. This indicates that the 
existence of the second center requires a very specific choice of the fluid's
equation of state and carefully chosen initial conditions.

\section{Summary and conclusions}
\label{sec:concl}

We have used analytical and numerical techniques to analyze the static and
spherically perfect fluid field equations of general relativity in the presence
of a cosmological constant. A positive cosmological term can be viewed as an
external force having the effect of pulling matter apart. Hence, one can
expect that the radial size of matter spheres is increased due to $\Lambda$.
This naturally yields questions regarding the physical picture applicable
to these solutions. It turns out that the effects of the cosmological constant
lead to various different configurations, many of which have not been discussed
previously. 

By using Gauss coordinates relative to the $r=\mbox{constant}$ hypersurfaces,
we analyzed geometrically the properties of the vanishing pressure
surface that determines the boundary of the perfect fluid sphere. 
In the absence of the cosmological constant, going outwards, the area
of the respective group orbits are always increasing close to the zero
pressure surface.
This situation changes drastically
if $\Lambda$ is allowed to be relatively large in comparison with the
matter density. It is possible for the pressure to vanish exactly at
the maximum of the group orbits or even vanish where the group orbits
are decreasing. In the first case one has to join on the Nariai solution to
get the metric $C^1$ at boundary. 
In the latter case one matches the part of the Schwarzschild-de Sitter
solution containing the black hole horizon and the singularity. 
This is in contrast to the the small $\Lambda$ situation where the
vacuum region contains the cosmological horizon. 
Lastly, one is lead to
ask whether the matter can occupy the whole spacetime resulting in two
regular centers corresponding to a fully generalized Einstein static universe
where neither the energy density nor the pressure are constant.

We showed that the Whittaker solution can have its vanishing pressure surface
where the group orbits are decreasing, however, a second center is not possible.
On the other hand, the Tolman $IV$ solution does allow for a second regular
center, a solution that might be named Tolman $IV$ Einstein universe.
By numerically integrating the field equations for physically motivated 
equations of state we showed that in general the pressure can vanish where
the group orbits are decreasing and consequently at the maximum for sufficiently
fine tuned initial conditions. We also obtained analytical bounds that
the cosmological constant has to satisfy to allow for such situations. However,
we were not able to show, that in general, solutions with two regular centers
for a given equation of state exist. This observation has analytical support since 
the conditions under which the solutions have two regular centers are very restrictive.
Note however, that the special Tolman $IV$ equation of state~(\ref{eom:tolman}) 
admits solutions with a second regular center.

Ever since the first exact matter solutions have been obtained, static and 
spherically symmetric perfect fluid spacetimes have remained a subject of great
interest. The presence of matter that effectively acts like a perfect fluid
with unusual equations of state, such as $p/\rho = -1$, drastically
changes the geometry of known solutions. 

\acknowledgments
We thank P\'{e}ter Forg\'{a}cs and Roy Maartens for the useful discussions.
GF has been supported by OTKA Grants No. K67942, K61636 and NI68228, 
and would like to thank the Bolyai Foundation for financial support.

\appendix

\section{Field equations and Riemann tensor}
\label{app}
The non-vanishing Riemann tensor components are
\begin{alignat}{2}
      R^{rt}{}_{rt} &= -\nu'^2-\nu'', &\qquad
      R^{\theta\phi}{}_{\theta\phi} &= \frac{1-R'^2}{R^2},
      \nonumber \\
      R^{r\theta}{}_{r\theta}= R^{r\phi}{}_{r\phi}&= -\frac{R''}{R},
      &\qquad
      R^{\theta t}{}_{\theta t}= R^{\phi t}{}_{\phi t}&= -\nu' \frac{R'}{R}.
      \label{riemann}
\end{alignat}
One can rewrite the field equations (\ref{an2})--(\ref{an4}) to get
\begin{align}
      \frac{R''}{R} &= -4\pi \rho -\frac{\Lambda}{3} + \frac{m}{R^3},
      \label{an2a} \\
      \nu'\frac{R'}{R} &= 4\pi p - \frac{\Lambda}{3} + \frac{m}{R^3},
      \label{an3a} \\
      \nu'^2 +\nu'' &= 4\pi(\rho+ p) - \frac{2m}{R^3} -\frac{\Lambda}{3},
      \label{an4a}
\end{align}
and hence the Riemann tensor in terms of physical quantities
\begin{align}
      R^{rt}{}_{rt} &= -4\pi(\rho+ p) + \frac{2m}{R^3} + \frac{\Lambda}{3}, \\
      R^{\theta\phi}{}_{\theta\phi} &= \frac{2m}{R^3} + \frac{\Lambda}{3}, \\ 
      R^{r\theta}{}_{r\theta} = R^{r\phi}{}_{r\phi}
      &= 4\pi \rho +\frac{\Lambda}{3} - \frac{m}{R^3},\\
      R^{\theta t}{}_{\theta t} = R^{\phi t}{}_{\phi t} 
      &= -4\pi p + \frac{\Lambda}{3} - \frac{m}{R^3}.
      \label{riemann1}
\end{align}


\begin{thebibliography}{46}
\expandafter\ifx\csname natexlab\endcsname\relax\def\natexlab#1{#1}\fi
\expandafter\ifx\csname bibnamefont\endcsname\relax
  \def\bibnamefont#1{#1}\fi
\expandafter\ifx\csname bibfnamefont\endcsname\relax
  \def\bibfnamefont#1{#1}\fi
\expandafter\ifx\csname citenamefont\endcsname\relax
  \def\citenamefont#1{#1}\fi
\expandafter\ifx\csname url\endcsname\relax
  \def\url#1{\texttt{#1}}\fi
\expandafter\ifx\csname urlprefix\endcsname\relax\def\urlprefix{URL }\fi
\providecommand{\bibinfo}[2]{#2}
\providecommand{\eprint}[2][]{\url{#2}}

\bibitem[{\citenamefont{Weyl}(1919)}]{Weyl:1919}
\bibinfo{author}{\bibfnamefont{H.}~\bibnamefont{Weyl}},
  \bibinfo{journal}{Physikalische Zeitschrift} \textbf{\bibinfo{volume}{20}},
  \bibinfo{pages}{31} (\bibinfo{year}{1919}).

\bibitem[{\citenamefont{B{\"o}hmer}(2002)}]{Boehmer:2002gg}
\bibinfo{author}{\bibfnamefont{C.~G.} \bibnamefont{B{\"o}hmer}}
  (\bibinfo{year}{2002}), \bibinfo{note}{{unpublished Diploma thesis}},
  \eprint{gr-qc/0308057}.

\bibitem[{\citenamefont{B{\"o}hmer}(2004{\natexlab{a}})}]{Boehmer:2003uz}
\bibinfo{author}{\bibfnamefont{C.~G.} \bibnamefont{B{\"o}hmer}},
  \bibinfo{journal}{Gen. Rel. Grav.} \textbf{\bibinfo{volume}{36}},
  \bibinfo{pages}{1039} (\bibinfo{year}{2004}{\natexlab{a}}),
  \eprint{gr-qc/0312027}.

\bibitem[{\citenamefont{Xu et~al.}(1986)\citenamefont{Xu, Wu, and
  Huang}}]{Xu:1986ka}
\bibinfo{author}{\bibfnamefont{C.-m.} \bibnamefont{Xu}},
  \bibinfo{author}{\bibfnamefont{X.-j.} \bibnamefont{Wu}}, \bibnamefont{and}
  \bibinfo{author}{\bibfnamefont{Z.}~\bibnamefont{Huang}}
  (\bibinfo{year}{1986}), \bibinfo{note}{iC-86-392}.

\bibitem[{\citenamefont{Stuchl{\'{\i}}k}(2000)}]{Stuchlik:2000}
\bibinfo{author}{\bibfnamefont{Z.}~\bibnamefont{Stuchl{\'{\i}}k}},
  \bibinfo{journal}{Acta Physica Slovaca} \textbf{\bibinfo{volume}{50}},
  \bibinfo{pages}{219} (\bibinfo{year}{2000}).

\bibitem[{\citenamefont{Nariai}(1999{\natexlab{a}})}]{Nariai1}
\bibinfo{author}{\bibfnamefont{H.}~\bibnamefont{Nariai}},
  \bibinfo{journal}{Gen. Rel. Grav.} \textbf{\bibinfo{volume}{31}},
  \bibinfo{pages}{963} (\bibinfo{year}{1999}{\natexlab{a}}),
  \bibinfo{note}{originally published in \emph{The Science Reports of the
  Tohoku University} Series I, vol. \textbf{XXXV}, No. 1 (1951), p. 46-57}.

\bibitem[{\citenamefont{Nariai}(1999{\natexlab{b}})}]{Nariai2}
\bibinfo{author}{\bibfnamefont{H.}~\bibnamefont{Nariai}},
  \bibinfo{journal}{Gen. Rel. Grav.} \textbf{\bibinfo{volume}{31}},
  \bibinfo{pages}{951} (\bibinfo{year}{1999}{\natexlab{b}}),
  \bibinfo{note}{originally published in \emph{The Science Reports of the
  Tohoku University} Series I, vol. \textbf{XXXIV}, No. 3 (1950), p. 160-167}.

\bibitem[{\citenamefont{Einstein}(1917)}]{Einstein:1917ce}
\bibinfo{author}{\bibfnamefont{A.}~\bibnamefont{Einstein}},
  \bibinfo{journal}{Sitzungsber. Preuss. Akad. Wiss. Berlin (Math. Phys. )} pp.
  \bibinfo{pages}{142--152} (\bibinfo{year}{1917}).

\bibitem[{\citenamefont{Ibrahim and Nutku}(1976)}]{Ibrahim:1976}
\bibinfo{author}{\bibfnamefont{A.}~\bibnamefont{Ibrahim}} \bibnamefont{and}
  \bibinfo{author}{\bibfnamefont{Y.}~\bibnamefont{Nutku}},
  \bibinfo{journal}{Gen. Rel. Grav.} \textbf{\bibinfo{volume}{7}},
  \bibinfo{pages}{949} (\bibinfo{year}{1976}).

\bibitem[{\citenamefont{Collins}(1988)}]{Collins:1988}
\bibinfo{author}{\bibfnamefont{C.~B.} \bibnamefont{Collins}},
  \bibinfo{journal}{Gen. Rel. Grav.} \textbf{\bibinfo{volume}{20}},
  \bibinfo{pages}{847} (\bibinfo{year}{1988}).

\bibitem[{\citenamefont{Gergely and Maartens}(2002)}]{Gergely:2001tn}
\bibinfo{author}{\bibfnamefont{L.~A.} \bibnamefont{Gergely}} \bibnamefont{and}
  \bibinfo{author}{\bibfnamefont{R.}~\bibnamefont{Maartens}},
  \bibinfo{journal}{Class. Quant. Grav.} \textbf{\bibinfo{volume}{19}},
  \bibinfo{pages}{213} (\bibinfo{year}{2002}), \eprint{gr-qc/0105058}.

\bibitem[{\citenamefont{B{\"o}hmer}(2004{\natexlab{b}})}]{Boehmer:2003iv}
\bibinfo{author}{\bibfnamefont{C.~G.} \bibnamefont{B{\"o}hmer}},
  \bibinfo{journal}{Class. Quant. Grav.} \textbf{\bibinfo{volume}{21}},
  \bibinfo{pages}{1119} (\bibinfo{year}{2004}{\natexlab{b}}),
  \eprint{gr-qc/0310058}.

\bibitem[{\citenamefont{B{\"o}hmer et~al.}(2007)\citenamefont{B{\"o}hmer,
  Hollenstein, and Lobo}}]{Boehmer:2007tr}
\bibinfo{author}{\bibfnamefont{C.~G.} \bibnamefont{B{\"o}hmer}},
  \bibinfo{author}{\bibfnamefont{L.}~\bibnamefont{Hollenstein}},
  \bibnamefont{and} \bibinfo{author}{\bibfnamefont{F.~S.~N.}
  \bibnamefont{Lobo}}, \bibinfo{journal}{Phys. Rev.}
  \textbf{\bibinfo{volume}{D76}}, \bibinfo{pages}{084005}
  (\bibinfo{year}{2007}), \eprint{arXiv:0706.1663 [gr-qc]}.

\bibitem[{\citenamefont{Parisi et~al.}(2007)\citenamefont{Parisi, Bruni,
  Maartens, and Vandersloot}}]{Parisi:2007kv}
\bibinfo{author}{\bibfnamefont{L.}~\bibnamefont{Parisi}},
  \bibinfo{author}{\bibfnamefont{M.}~\bibnamefont{Bruni}},
  \bibinfo{author}{\bibfnamefont{R.}~\bibnamefont{Maartens}}, \bibnamefont{and}
  \bibinfo{author}{\bibfnamefont{K.}~\bibnamefont{Vandersloot}},
  \bibinfo{journal}{Class. Quant. Grav.} \textbf{\bibinfo{volume}{24}},
  \bibinfo{pages}{6243} (\bibinfo{year}{2007}), \eprint{arXiv:0706.4431
  [gr-qc]}.

\bibitem[{\citenamefont{Chodos et~al.}(1974{\natexlab{a}})\citenamefont{Chodos,
  Jaffe, Johnson, Thorn, and Weisskopf}}]{Chodos:1974je}
\bibinfo{author}{\bibfnamefont{A.}~\bibnamefont{Chodos}},
  \bibinfo{author}{\bibfnamefont{R.~L.} \bibnamefont{Jaffe}},
  \bibinfo{author}{\bibfnamefont{K.}~\bibnamefont{Johnson}},
  \bibinfo{author}{\bibfnamefont{C.~B.} \bibnamefont{Thorn}}, \bibnamefont{and}
  \bibinfo{author}{\bibfnamefont{V.~F.} \bibnamefont{Weisskopf}},
  \bibinfo{journal}{Phys. Rev.} \textbf{\bibinfo{volume}{D9}},
  \bibinfo{pages}{3471} (\bibinfo{year}{1974}{\natexlab{a}}).

\bibitem[{\citenamefont{Chodos et~al.}(1974{\natexlab{b}})\citenamefont{Chodos,
  Jaffe, Johnson, and Thorn}}]{Chodos:1974pn}
\bibinfo{author}{\bibfnamefont{A.}~\bibnamefont{Chodos}},
  \bibinfo{author}{\bibfnamefont{R.~L.} \bibnamefont{Jaffe}},
  \bibinfo{author}{\bibfnamefont{K.}~\bibnamefont{Johnson}}, \bibnamefont{and}
  \bibinfo{author}{\bibfnamefont{C.~B.} \bibnamefont{Thorn}},
  \bibinfo{journal}{Phys. Rev.} \textbf{\bibinfo{volume}{D10}},
  \bibinfo{pages}{2599} (\bibinfo{year}{1974}{\natexlab{b}}).

\bibitem[{\citenamefont{DeGrand et~al.}(1975)\citenamefont{DeGrand, Jaffe,
  Johnson, and Kiskis}}]{DeGrand:1975cf}
\bibinfo{author}{\bibfnamefont{T.~A.} \bibnamefont{DeGrand}},
  \bibinfo{author}{\bibfnamefont{R.~L.} \bibnamefont{Jaffe}},
  \bibinfo{author}{\bibfnamefont{K.}~\bibnamefont{Johnson}}, \bibnamefont{and}
  \bibinfo{author}{\bibfnamefont{J.~E.} \bibnamefont{Kiskis}},
  \bibinfo{journal}{Phys. Rev.} \textbf{\bibinfo{volume}{D12}},
  \bibinfo{pages}{2060} (\bibinfo{year}{1975}).

\bibitem[{\citenamefont{Kaup}(1968)}]{Kaup:1968}
\bibinfo{author}{\bibfnamefont{D.~J.} \bibnamefont{Kaup}},
  \bibinfo{journal}{Phys. Rev.} \textbf{\bibinfo{volume}{172}},
  \bibinfo{pages}{1331} (\bibinfo{year}{1968}).

\bibitem[{\citenamefont{Ruffini and Bonazzola}(1969)}]{Ruffini:1969qy}
\bibinfo{author}{\bibfnamefont{R.}~\bibnamefont{Ruffini}} \bibnamefont{and}
  \bibinfo{author}{\bibfnamefont{S.}~\bibnamefont{Bonazzola}},
  \bibinfo{journal}{Phys. Rev.} \textbf{\bibinfo{volume}{187}},
  \bibinfo{pages}{1767} (\bibinfo{year}{1969}).

\bibitem[{\citenamefont{B{\"o}hmer and Vandersloot}(2007)}]{Boehmer:2007wi}
\bibinfo{author}{\bibfnamefont{C.~G.} \bibnamefont{B{\"o}hmer}}
  \bibnamefont{and}
  \bibinfo{author}{\bibfnamefont{K.}~\bibnamefont{Vandersloot}},
  \bibinfo{journal}{Phys. Rev.} \textbf{\bibinfo{volume}{D76}},
  \bibinfo{pages}{104030} (\bibinfo{year}{2007}), \eprint{arXiv:0709.2129
  [gr-qc]}.

\bibitem[{\citenamefont{B{\"o}hmer and Harko}(2005)}]{Boehmer:2005kk}
\bibinfo{author}{\bibfnamefont{C.~G.} \bibnamefont{B{\"o}hmer}}
  \bibnamefont{and} \bibinfo{author}{\bibfnamefont{T.}~\bibnamefont{Harko}},
  \bibinfo{journal}{Phys. Rev.} \textbf{\bibinfo{volume}{D71}},
  \bibinfo{pages}{084026} (\bibinfo{year}{2005}), \eprint{gr-qc/0504075}.

\bibitem[{\citenamefont{Hledik et~al.}(2007)\citenamefont{Hledik, Stuchlik, and
  Mrazova}}]{Hledik:2007tr}
\bibinfo{author}{\bibfnamefont{S.}~\bibnamefont{Hledik}},
  \bibinfo{author}{\bibfnamefont{Z.}~\bibnamefont{Stuchlik}}, \bibnamefont{and}
  \bibinfo{author}{\bibfnamefont{K.}~\bibnamefont{Mrazova}}
  (\bibinfo{year}{2007}), \eprint{gr-qc/0701051}.

\bibitem[{\citenamefont{Iorio}(2006)}]{Iorio:2005vw}
\bibinfo{author}{\bibfnamefont{L.}~\bibnamefont{Iorio}}, \bibinfo{journal}{Int.
  J. Mod. Phys.} \textbf{\bibinfo{volume}{D15}}, \bibinfo{pages}{473}
  (\bibinfo{year}{2006}), \eprint{gr-qc/0511137}.

\bibitem[{\citenamefont{Sereno and Jetzer}(2006)}]{Sereno:2006re}
\bibinfo{author}{\bibfnamefont{M.}~\bibnamefont{Sereno}} \bibnamefont{and}
  \bibinfo{author}{\bibfnamefont{P.}~\bibnamefont{Jetzer}},
  \bibinfo{journal}{Phys. Rev.} \textbf{\bibinfo{volume}{D73}},
  \bibinfo{pages}{063004} (\bibinfo{year}{2006}), \eprint{astro-ph/0602438}.

\bibitem[{\citenamefont{Jetzer and Sereno}(2006)}]{Jetzer:2006gn}
\bibinfo{author}{\bibfnamefont{P.}~\bibnamefont{Jetzer}} \bibnamefont{and}
  \bibinfo{author}{\bibfnamefont{M.}~\bibnamefont{Sereno}},
  \bibinfo{journal}{Phys. Rev.} \textbf{\bibinfo{volume}{D73}},
  \bibinfo{pages}{044015} (\bibinfo{year}{2006}), \eprint{astro-ph/0601612}.

\bibitem[{\citenamefont{Kagramanova et~al.}(2006)\citenamefont{Kagramanova,
  Kunz, and Lammerzahl}}]{Kagramanova:2006ax}
\bibinfo{author}{\bibfnamefont{V.}~\bibnamefont{Kagramanova}},
  \bibinfo{author}{\bibfnamefont{J.}~\bibnamefont{Kunz}}, \bibnamefont{and}
  \bibinfo{author}{\bibfnamefont{C.}~\bibnamefont{Lammerzahl}},
  \bibinfo{journal}{Phys. Lett.} \textbf{\bibinfo{volume}{B634}},
  \bibinfo{pages}{465} (\bibinfo{year}{2006}), \eprint{gr-qc/0602002}.

\bibitem[{\citenamefont{Balaguera-Antol{\'{\i}}nez
  et~al.}(2005)\citenamefont{Balaguera-Antol{\'{\i}}nez, B{\"o}hmer, and
  Nowakowski}}]{Balaguera-Antolinez:2004sg}
\bibinfo{author}{\bibfnamefont{A.}~\bibnamefont{Balaguera-Antol{\'{\i}}nez}},
  \bibinfo{author}{\bibfnamefont{C.~G.} \bibnamefont{B{\"o}hmer}},
  \bibnamefont{and}
  \bibinfo{author}{\bibfnamefont{M.}~\bibnamefont{Nowakowski}},
  \bibinfo{journal}{Int. J. Mod. Phys.} \textbf{\bibinfo{volume}{D14}},
  \bibinfo{pages}{1507} (\bibinfo{year}{2005}), \eprint{gr-qc/0409004}.

\bibitem[{\citenamefont{Balaguera-Antol{\'{\i}}nez
  et~al.}(2006)\citenamefont{Balaguera-Antol{\'{\i}}nez, B{\"o}hmer, and
  Nowakowski}}]{Balaguera-Antolinez:2005wg}
\bibinfo{author}{\bibfnamefont{A.}~\bibnamefont{Balaguera-Antol{\'{\i}}nez}},
  \bibinfo{author}{\bibfnamefont{C.~G.} \bibnamefont{B{\"o}hmer}},
  \bibnamefont{and}
  \bibinfo{author}{\bibfnamefont{M.}~\bibnamefont{Nowakowski}},
  \bibinfo{journal}{Class. Quant. Grav.} \textbf{\bibinfo{volume}{23}},
  \bibinfo{pages}{485} (\bibinfo{year}{2006}), \eprint{gr-qc/0511057}.

\bibitem[{\citenamefont{Balaguera-Antol{\'{\i}}nez and
  Nowakowski}(2005)}]{Balaguera-Antolinez:2005yy}
\bibinfo{author}{\bibfnamefont{A.}~\bibnamefont{Balaguera-Antol{\'{\i}}nez}}
  \bibnamefont{and}
  \bibinfo{author}{\bibfnamefont{M.}~\bibnamefont{Nowakowski}},
  \bibinfo{journal}{Astron. Astrophys.} \textbf{\bibinfo{volume}{441}},
  \bibinfo{pages}{23} (\bibinfo{year}{2005}), \eprint{astro-ph/0511738}.

\bibitem[{\citenamefont{Balaguera-Antol{\'{\i}}nez
  et~al.}(2007)\citenamefont{Balaguera-Antol{\'{\i}}nez, Mota, and
  Nowakowski}}]{BalagueraAntolinez:2007sf}
\bibinfo{author}{\bibfnamefont{A.}~\bibnamefont{Balaguera-Antol{\'{\i}}nez}},
  \bibinfo{author}{\bibfnamefont{D.~F.} \bibnamefont{Mota}}, \bibnamefont{and}
  \bibinfo{author}{\bibfnamefont{M.}~\bibnamefont{Nowakowski}}
  (\bibinfo{year}{2007}), \eprint{arXiv:0708.2980 [astro-ph]}.

\bibitem[{\citenamefont{Balaguera-Antol{\'{\i}}nez and
  Nowakowski}(2007)}]{Balaguera-Antolinez:2007mx}
\bibinfo{author}{\bibfnamefont{A.}~\bibnamefont{Balaguera-Antol{\'{\i}}nez}}
  \bibnamefont{and}
  \bibinfo{author}{\bibfnamefont{M.}~\bibnamefont{Nowakowski}},
  \bibinfo{journal}{Class. Quant. Grav.} \textbf{\bibinfo{volume}{24}},
  \bibinfo{pages}{2677} (\bibinfo{year}{2007}), \eprint{arXiv:0704.1871
  [gr-qc]}.

\bibitem[{\citenamefont{Rindler and Ishak}(2007)}]{Rindler:2007zz}
\bibinfo{author}{\bibfnamefont{W.}~\bibnamefont{Rindler}} \bibnamefont{and}
  \bibinfo{author}{\bibfnamefont{M.}~\bibnamefont{Ishak}},
  \bibinfo{journal}{Phys. Rev.} \textbf{\bibinfo{volume}{D76}},
  \bibinfo{pages}{043006} (\bibinfo{year}{2007}), \eprint{arXiv:0709.2948
  [astro-ph]}.

\bibitem[{\citenamefont{Sereno}(2007)}]{Sereno:2007rm}
\bibinfo{author}{\bibfnamefont{M.}~\bibnamefont{Sereno}}
  (\bibinfo{year}{2007}), \eprint{arXiv:0711.1802 [astro-ph]}.

\bibitem[{\citenamefont{Whittaker}(1968)}]{Whittaker:1968}
\bibinfo{author}{\bibfnamefont{J.~M.} \bibnamefont{Whittaker}},
  \bibinfo{journal}{Proc. Roy. Soc.} \textbf{\bibinfo{volume}{A306}},
  \bibinfo{pages}{1} (\bibinfo{year}{1968}).

\bibitem[{\citenamefont{Tolman}(1939)}]{Tolman:1939jz}
\bibinfo{author}{\bibfnamefont{R.~C.} \bibnamefont{Tolman}},
  \bibinfo{journal}{Phys. Rev.} \textbf{\bibinfo{volume}{55}},
  \bibinfo{pages}{364} (\bibinfo{year}{1939}).

\bibitem[{\citenamefont{Wahlquist}(1968)}]{Wahlquist:1968}
\bibinfo{author}{\bibfnamefont{H.~D.} \bibnamefont{Wahlquist}},
  \bibinfo{journal}{Phys. Rev.} \textbf{\bibinfo{volume}{172}},
  \bibinfo{pages}{1291} (\bibinfo{year}{1968}).

\bibitem[{\citenamefont{Darmois}(1927)}]{Darmois:1927}
\bibinfo{author}{\bibfnamefont{G.}~\bibnamefont{Darmois}},
  \emph{\bibinfo{title}{M{\'e}morial de Sciences Math{\'e}matiques}}, vol.
  \bibinfo{volume}{XXV} (\bibinfo{publisher}{Gauthier-Villars, Paris},
  \bibinfo{year}{1927}).

\bibitem[{\citenamefont{Israel}(1966)}]{Israel:1966rt}
\bibinfo{author}{\bibfnamefont{W.}~\bibnamefont{Israel}},
  \bibinfo{journal}{Nuovo Cim.} \textbf{\bibinfo{volume}{B44S10}},
  \bibinfo{pages}{1} (\bibinfo{year}{1966}).

\bibitem[{\citenamefont{Zannias}(1990)}]{Zannias:1990}
\bibinfo{author}{\bibfnamefont{T.}~\bibnamefont{Zannias}},
  \bibinfo{journal}{Phys. Rev.} \textbf{\bibinfo{volume}{D41}},
  \bibinfo{pages}{3252} (\bibinfo{year}{1990}).

\bibitem[{\citenamefont{Delgaty and Lake}(1998)}]{Delgaty:1998uy}
\bibinfo{author}{\bibfnamefont{M.~S.~R.} \bibnamefont{Delgaty}}
  \bibnamefont{and} \bibinfo{author}{\bibfnamefont{K.}~\bibnamefont{Lake}},
  \bibinfo{journal}{Comput. Phys. Commun.} \textbf{\bibinfo{volume}{115}},
  \bibinfo{pages}{395} (\bibinfo{year}{1998}), \eprint{gr-qc/9809013}.

\bibitem[{\citenamefont{Rendall and Schmidt}(1991)}]{Rendall:1991hg}
\bibinfo{author}{\bibfnamefont{A.~D.} \bibnamefont{Rendall}} \bibnamefont{and}
  \bibinfo{author}{\bibfnamefont{B.~G.} \bibnamefont{Schmidt}},
  \bibinfo{journal}{Class. Quant. Grav.} \textbf{\bibinfo{volume}{8}},
  \bibinfo{pages}{985} (\bibinfo{year}{1991}).

\bibitem[{\citenamefont{Baumgarte and Rendall}(1993)}]{Baumgarte:1993}
\bibinfo{author}{\bibfnamefont{T.~W.} \bibnamefont{Baumgarte}}
  \bibnamefont{and} \bibinfo{author}{\bibfnamefont{A.~D.}
  \bibnamefont{Rendall}}, \bibinfo{journal}{Class. Quant. Grav.}
  \textbf{\bibinfo{volume}{10}}, \bibinfo{pages}{327} (\bibinfo{year}{1993}).

\bibitem[{\citenamefont{Mars et~al.}(1996)\citenamefont{Mars, Martin-Prats, and
  Senovilla}}]{Mars:1996gd}
\bibinfo{author}{\bibfnamefont{M.}~\bibnamefont{Mars}},
  \bibinfo{author}{\bibfnamefont{M.~M.} \bibnamefont{Martin-Prats}},
  \bibnamefont{and} \bibinfo{author}{\bibfnamefont{J.~M.~M.}
  \bibnamefont{Senovilla}}, \bibinfo{journal}{Phys. Lett.}
  \textbf{\bibinfo{volume}{A218}}, \bibinfo{pages}{147} (\bibinfo{year}{1996}),
  \eprint{gr-qc/0202003}.

\bibitem[{\citenamefont{B{\"o}hmer}(2005)}]{Boehmer:2004nu}
\bibinfo{author}{\bibfnamefont{C.~G.} \bibnamefont{B{\"o}hmer}},
  \bibinfo{journal}{Ukr. J. Phys.} \textbf{\bibinfo{volume}{50}},
  \bibinfo{pages}{1219} (\bibinfo{year}{2005}), \eprint{gr-qc/0409030}.

\bibitem[{\citenamefont{Oppenheimer and Volkoff}(1939)}]{Oppenheimer:1939ne}
\bibinfo{author}{\bibfnamefont{J.~R.} \bibnamefont{Oppenheimer}}
  \bibnamefont{and} \bibinfo{author}{\bibfnamefont{G.~M.}
  \bibnamefont{Volkoff}}, \bibinfo{journal}{Phys. Rev.}
  \textbf{\bibinfo{volume}{55}}, \bibinfo{pages}{374} (\bibinfo{year}{1939}).

\bibitem[{\citenamefont{Fodor}(2000)}]{Fodor:2000gu}
\bibinfo{author}{\bibfnamefont{G.}~\bibnamefont{Fodor}} (\bibinfo{year}{2000}),
  \eprint{gr-qc/0011040}.

\end{thebibliography}
\end{document}